%% file: sample-acmsmall.tex
\documentclass[acmsmall]{acmart}

\input{packages.tex}
\input{variables-and-settings.tex}

\AtBeginDocument{%
  }

\begin{document}


\title[Cartographers in Cubicles]{Cartographers in Cubicles: How Training and Preferences of Mapmakers Interplay with Structures and Norms in Not-for-Profit Organizations}

\author{Arpit Narechania}
\email{arpit@ust.hk}
\orcid{0000-0001-6980-3686}
\affiliation{%
    \institution{The Hong Kong University of Science and Technology}
  \country{Hong Kong SAR, China}
}
\authornote{This work was done in part when the author was affiliated with Georgia Institute of Technology.}

\author{Alex Endert}
\orcid{0000-0002-6914-610X}
\affiliation{%
  \institution{Georgia Institute of Technology}
  \city{Atlanta}
  \state{Georgia}
  \country{USA}
}
\email{endert@gatech.edu}

\author{Clio Andris}
\orcid{0000-0002-8559-5079}
\affiliation{%
  \institution{Georgia Institute of Technology}
  \city{Atlanta}
  \state{Georgia}
  \country{USA}
}
\email{clio@gatech.edu}

\renewcommand{\shortauthors}{Narechania, Endert, and Andris}

\begin{abstract}
  Choropleth maps are a common and effective way to visualize geographic thematic data. Although cartographers have established many principles about map design, data binning and color usage, less is known about how mapmakers make individual decisions in practice. We interview 16 cartographers and geographic information systems (GIS) experts from 13 government organizations, NGOs, and federal agencies about their choropleth mapmaking decisions and workflows. We categorize our findings and report on how mapmakers follow cartographic guidelines and personal rules of thumb, collaborate with other stakeholders within and outside their organization, and how organizational structures and norms are tied to decision-making during data preparation, data analysis, data binning, map styling, and map post-processing. We find several points of variation as well as regularity across mapmakers and organizations and present takeaways to inform cartographic education and practice, including broader implications and opportunities for CSCW, HCI, and information visualization researchers and practitioners.
\end{abstract}

\ccsdesc[500]{Human-centered computing~Geographic visualization}
\ccsdesc[500]{Human-centered computing~Empirical studies in collaborative and social computing}
\ccsdesc[500]{Human-centered computing~Visualization theory, concepts and paradigms}

\keywords{cartography, geographic information science, choropleth map, visualization, interview study, qualitative research, collaboration, organization}

\begin{teaserfigure}
  \centering
  \includegraphics[width=0.95\textwidth]{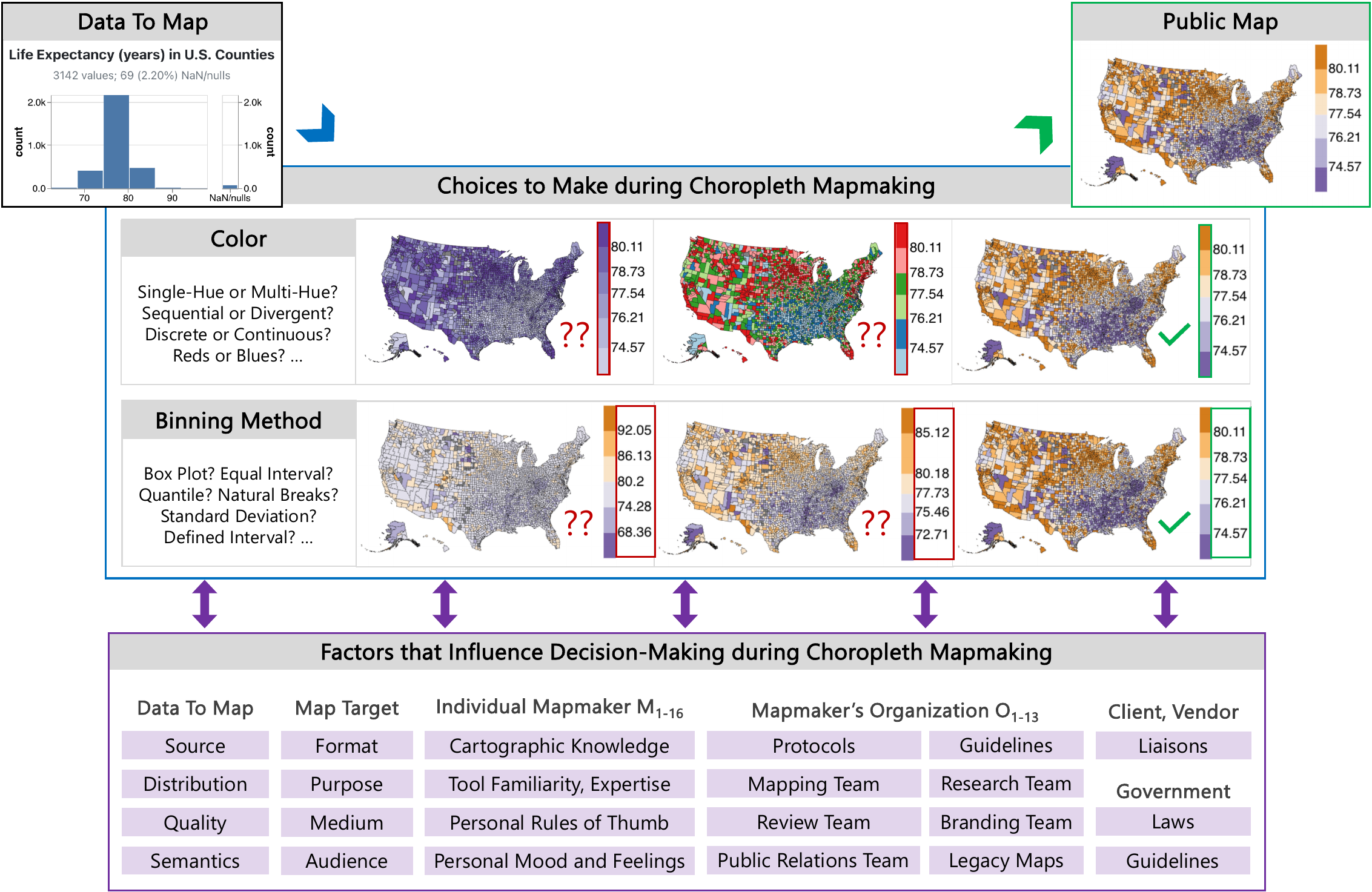}
  \caption{Factors that influence decision-making during choropleth mapmaking, as derived from our interview study.}
  \label{fig:teaser}
  \Description{This teaser figure shows four blocks, linked to each other. First, "Data to Map" showing the data profile and a histogram visualization of the data distribution for a Life Expectancy (in years) across U.S. counties dataset; second, "Choices to make during Choropleth Mapmaking" comprising two high-level categories: Colors and Binning Methods with small multiples showing how a choropleth can narrate different stories with different color schemes and binning methods, respectively; third, "Factors that Influence Decision-Making during Choropleth Mapmaking" comprising high-level category blocks in Data To Map, Map Target, Individual Mapmaker, Mapmaker's Organization, Client, Vendor, and Government; fourth: the "Public Map" that is output as an artifact of the decision-making.}
\end{teaserfigure}


\maketitle

\section{Introduction}
\label{section:introduction}
\input{src/sections/01-introduction.tex}

\section{Related Work}
\label{section:relatedwork}
\input{src/sections/02-relatedwork.tex}

\section{Interviews with Cartographers and GIS Experts}
\label{section:expert-interviews}
\input{src/sections/03-expert-interviews.tex}

\section{Choropleth Mapmaking in Government Organizations and NGO\MakeLowercase{s}}
\label{section:expert-interviews-findings}
\input{src/sections/04-expert-interviews-findings.tex}

\section{Discussion: Broader Implications and Opportunities for CSCW, HCI, and information visualization}
\label{section:discussion}
\input{src/sections/06-discussion.tex}

\section{Limitations and Future Work}
\label{section:limitations}
\input{src/sections/07-limitations.tex}

%
\begin{acks}
This material is based upon work supported by NSF IIS-1750474. We are grateful to the cartographers and GIS experts for interviewing with us. We also thank the Georgia Tech Visualization Lab, Karthik Bhat, and anonymous reviewers for their feedback at different stages of this work. We used Google Scholar~\cite{googlescholar}, vitaLITy~\cite{vitality2022narechania}, and vitaLITy~2~\cite{an2024vitality2} to assist us during our literature review.
\end{acks}

\bibliographystyle{ACM-Reference-Format}
\bibliography{sample-base}


\end{document}

%% file: packages.tex
\usepackage{microtype}                 
\PassOptionsToPackage{warn}{textcomp}  
\usepackage{textcomp}                  
\usepackage{mathptmx}                  
\usepackage{times}                     
\usepackage{tabu}                      
\usepackage{booktabs}                  
\usepackage{listings}
\usepackage{enumitem}
\usepackage{multirow}
\usepackage{tabularx}
\usepackage{xcolor}
\usepackage{xspace}
\usepackage{amsmath}
\usepackage{color}
\usepackage{soul}
\usepackage{graphicx}
\usepackage{caption}

%% file: variables-and-settings.tex
\definecolor{orange}{RGB}{237,125,49}
\definecolor{custompurple}{RGB}{104,50,154}
\definecolor{pink}{RGB}{245,114,255}
\definecolor{black}{RGB}{0,0,0}
\definecolor{TEAL}{RGB}{0,128,128}
\definecolor{BLUE}{RGB}{0,0,255}
\definecolor{RED}{RGB}{255,0,000}
\definecolor{cblue}{RGB}{0,112,192}

\newcommand{\expert}[1]{\textcolor{teal}{$M_{#1}$}}
\newcommand{\org}[1]{\textcolor{custompurple}{$O_{#1}$}}

\newcommand{\paragraphHeadingSpace}{\vspace{4px}}
\newcommand{\bpstart}[1]{
\noindent{\textbf{#1}}%
}

\newcommand{\cut}[1]{}

%% file: src/sections/01-introduction.tex
A choropleth map is a type of thematic map~\cite{bartz1979place} where administrative units or geographic regions are filled with different colors, shades, or patterns to represent the magnitude of an attribute associated with the underlying geography. 
These maps represent spatial patterns and variations in data and facilitate comparisons across geographies based on data~\cite{slocum2022thematic}.
These maps also allow for pre-attentive processing and do not require advanced language skills, which makes them relatable and easy to understand~\cite{slocum2022thematic}.
They are commonly used to visualize economic~\cite{ver2013food}, demographic~\cite{cnn2020uselectionchoropleth}, and public health~\cite{cdc2022adultobesitychoropleth} data, and can be found in static~\cite{cdc2022adultobesitychoropleth}, interactive~\cite{cnn2020uselectionchoropleth}, and immersive~\cite{yang2020tilt} formats across print and digital media. One example of an impactful thematic map is the number of votes for a U.S. presidential candidate by county, which is used to communicate the preferences of voters in different parts of the country~\cite{nytimes2021uselectionchoropleth}. 

To make choropleth maps, cartographers and geographic information science (GIS) experts (henceforth referred to collectively as mapmakers) use established cartographic guidelines to choose appropriate data binning methods~\cite{jenks1963class}, color ramps or color spectra~\cite{brewer1994color, brewer1997mapping}, normalization strategies, map scales, and map projections~\cite{kessler2019working}. 
Their stylistic choices affect how map information is communicated, and certain choices can misrepresent the data~\cite{monmonier2018lie, openshaw1983modifiable}.

While much is known about map design principles, less is known about how professional mapmakers use (or amend) these principles when designing maps for the public.
For instance, what tools do they use? How do they determine binning methods and color palettes? 
Do they collaborate with other people?
How do their organizational structures and protocols impact or influence these decisions? 
What difficulties do they face during the process?
Documenting these decisions and practices can help identify challenges and perhaps inform future cartographic tools and workflows.

In the computer-supported cooperative work (CSCW), human-computer interaction (HCI), and information visualization literature, there have been many interview studies that have contributed to our understanding of collaborations within organizations~\cite{kandel2012enterprise, muller2019data}, the role of visualizations in practice~\cite{madanagopal2019analytic, newburger2023visualization}, and the design of visualizations~\cite{bako2022understanding, zhang2022visualization}.

In particular, CSCW has advanced our understanding of how technology mediates workers' practices in varied fields which is especially helpful in addressing potential exploitation arising from the use of new computing technologies. For instance, in healthcare, clinicians and administrative staff navigate the sociotechnical challenges of electronic medical record (EMR) systems, where organizational readiness, user training, and system usability influence their ability to adopt and effectively use these tools~\cite{cucciniello2015understanding}.
In education, teachers turned to digital platforms for remote teaching during the COVID-19 pandemic, and navigated challenges such as mitigating technological inequity (i.e., the digital divide) and adapting traditional in-person pedagogy techniques and materials to digital formats~\cite{hodges2020difference}. 
In platform-based labor, `gig' workers experience algorithmic management techniques that are designed to balance worker autonomy with maximizing profit at the expense of the worker~\cite{wood2019good}.
Some of these fields also use spatially-explicit technologies (such as map technologies). 
We argue that CSCW can include more research on spatial technologies and mapmaking, and such research could benefit from adopting frameworks, research questions, and findings from the CSCW community.

In this article, we extend our knowledge by exploring how mapmaking training and education translates into practice and how organizational structures, processes, and protocols, and personal preferences impact decision-making in mapmaking. In particular, we find that mapmaking is an underexplored aspect of CSCW and HCI, despite the widespread use of thematic maps.
We interviewed 16 mapmakers from 13 globally-serving government organizations and NGOs (similar to the World Health Organization (WHO)) and federal agencies (similar to the U.S. Department of Agriculture).
We report on and characterize their choropleth mapmaking workflows (e.g., tools and techniques used) to prepare, analyze, and map data and also discuss the factors that influence decision-making during this process (Figure~\ref{fig:teaser}).
We chose these organizations as they are authoritative experts with a global reach and influence on societal development and public well-being. 
Their mapping efforts directly shape public policies, regulations, and international agreements, influencing areas such as 
public health~\cite{who_covid_map}, 
environmental protection~\cite{ipcc_climate_maps}, 
disaster management~\cite{fema_flood_maps}, 
human rights violations~\cite{amnesty_conflict_maps}, 
crime prevention~\cite{crime_mapping}, and 
global security~\cite{amnesty_violence_maps}, thereby also making this an important topic of study.

We find that mapmaking is a complex undertaking involving interdisciplinary collaborations within and across teams as well as organizations, wherein decisions are based on the data to be mapped, map target, mapmakers, subject matter experts, public relations experts, and organizational and government directives (Figure~\ref{fig:teaser}).
Notably, we find several points of variation as well as points of regularity across mapmakers.
For instance, choosing colors for a colorblind audience is a relatively universal consideration but can conflict with an organizational style guide or marketing-related constraints.
Similarly, choosing appropriate binning methods is challenging, and the eventual choices often vary across mapmakers, but it is commonly agreed upon that the resultant bins should be concise and easy to understand.
We present these as takeaways to inform cartographic education and practice by giving future mapmakers a sense of cartography in the workplace.
This work also contributes to CSCW and HCI scholarship by shedding light on how practitioners adhere to or adapt `textbook' principles when faced with production challenges. Moreover, it underscores the interplay between organizational structure, group dynamics, and their impact on both the process and outcomes, including how technology shapes and is shaped by collaborative work environments. 
Our findings also take readers behind-the-scenes to show how the graphics we consume, such as high-impact maps, are the results of nuanced and often-subjective decisions, including the surprising factor of mapmakers' `mood' on the day.

%% file: src/sections/02-relatedwork.tex
\subsection{Background on GIS and Cartographic Workflows, Principles, and Tools}
\label{sec:principles}
The process of map production uses project management paradigms, notably iterative communication between the map creators and the recipients~\cite{buckinham2019map}. 
Producing a map involves various steps including choosing the scale and scope, acquiring data, choosing stylistic elements through a `specification sheet', versioning, finalization (such as working with a graphic designer to enhance the map), and sharing digital files~\cite{buckinham2019map,slocum2022thematic}. 
There exist a variety of GIS and cartographic tools and libraries that help facilitate these steps. 
For instance, Esri products~\cite{esri} such as ArcPro and ArcGIS Online are widely used commercial software for geospatial data preparation, analysis, visualization, and management.
QGIS~\cite{qgis} is an open-source alternative to Esri products, and is popular with smaller organizations, academic institutions, and independent researchers.
Next, servers such as GeoServer help facilitate geospatial data and map sharing~\cite{geoserver}.
Additionally, GIS libraries such as ggplot2~\cite{ggplot2} offer customization and automation options, while styling tools such as ColorBrewer~\cite{harrower2003colorbrewer} enhance visual appeal. 
These tools collectively enable mapmakers to create maps that are informative, aesthetically pleasing, and tailored to project needs, available resources, and desired customization levels~\cite{buckinham2019map}.

In addition to teaching the process of mapmaking, cartographic training also encourages mapmakers to reflect on their (mapmaking) choices~\cite{howarth2020lesson}. There are many established cartographic principles and guidelines for creating maps, often derived from empirical studies.
For example, there is a rule of thumb that binning methods for choropleth maps should match the underlying data and should effectively communicate the data to the target audience~\cite{jenks1963class, tobler1973choropleth, brewer2002evaluation}. 
More specific guidance suggests that data bins for these (choropleth) maps range from three to seven~\cite{mersey1990colour, declerq1995choropleth, axismaps, brewer2002evaluation}.
Findings from our interview study shed light on the current status of organizational mapmaking workflows, principles, and tools. \\

\subsection{Collaboration within and across Organizations}
Effective collaboration between individuals and technology is fundamental for organizational success~\cite{grudin1994computer}. 
There exist many CSCW theories on collaboration~\cite{giddens1984constitution, star1989institutional, wasserman1994social, engestrom1999activity, latour2007reassembling} that describe the division of labor and communication patterns as well as the role of technology and organizational structures in cooperative work. 
For example, Activity Theory emphasizes the interplay between individual actions and the broader socio-cultural context~\cite{engestrom1999activity}, which helps contextualize a media creator's role in creating and educating the public. 
Social network analysis focuses on capturing relationships and information flow within networks~\cite{wasserman1994social}, which is useful for situating players in the mapmaking and sharing process. Similarly, Actor-Network Theory extends the analysis to include both human and non-human actors in shaping collaborative endeavors~\cite{latour2007reassembling}. 
Structuration Theory examines the recursive relationship between human actions and organizational structures~\cite{giddens1984constitution}, which helps explain the iteration involved in versioning maps and conducting interactive map critiques, etc. 
Finally, Boundary Object Theory highlights the role of artifacts and objects in facilitating collaboration across different social domains~\cite{star1989institutional}, which supports the narrative of mapmaking by centering the map as the media that encourages or enables collaboration.

Our work builds on prior studies illustrating these theories. For example, 
Kristoffersen~\cite{kristoffersen1999empirical} studied how people generally collaborate as part of their daily work activities,
Obermeyer~et~al.~\cite{obermeyer1990bureaucratic} highlighted bureaucratic barriers to innovation~\cite{obermeyer1990bureaucratic}, Orlikowski~\cite{orlikowski1992learning} discussed technology's dual role in enabling and constraining organizational contexts,
Campbell and Masser~\cite{campbell2020gis} described how a new (GIS) system is integrated into organizational workflows, 
Grudin~\cite{grudin1988groupware} analyzed groupware adoption challenges, emphasizing user-friendly design and incentive alignment for successful implementation, 
and Borgman~\cite{borgman2012s} discussed data management challenges in interdisciplinary organizations.

In general, the interdisciplinary nature of modern organizations emphasizes the need to consider differing perceptions and use-cases of data as well as the associated management challenges~\cite{borgman2012s}. 
Similarly, the global presence of modern organizations necessitates remote collaboration systems that support diverse work practices and sociotechnical environments without disrupting existing workflows~\cite{olson2000distance, erickson2000social}.
Agile collaboration is also crucial in times of crises, requiring improvisation and adaptive collaborative technologies to enable swift and coordinated responses in high-pressure situations~\cite{ley2014information}. 
In this work, we interview mapmakers from government organizations, NGOs, and federal agencies to understand how they collaboratively make maps, and put it in context with existing CSCW and HCI theories.

\subsection{Decision-Making within Organizations}
\label{subsection:decisionmakinginorgs}
Decision-making is central to the functioning of organizations, shaping their goals, operations, and outputs. Organizational decision-making has been extensively studied in literature, with theories focusing on individual, group, and systemic processes~\cite{march1993organizations, simon2013administrative, cyert2015behavioral}. For instance, theories in bounded rationality~\cite{simon1972theories} emphasize the limits of human cognition in decision-making within organizations~\cite{simon2013administrative}, and provide a framework for understanding how actors manage the complexity of organizational goals, including collaborative mapmaking.
Next, \emph{group} decision-making theories focus on the dynamics of team-based decisions, highlighting challenges like groupthink, polarization, and coordination~\cite{janis2008groupthink, sunstein1999law}, including territoriality~\cite{thom2009s} and competition~\cite{thom2010you}. For instance, behavioral theories proposed by Cyert and March emphasize the negotiation and satisficing (settling for a satisfactory solution rather than an optimal one) behaviors that are inherent to organizational environments~\cite{cyert2015behavioral}; these are relevant in scenarios where mapmakers must align their contributions with organizational or community goals.
Lastly, prior work on the influence of an \emph{individual}'s mood and emotional states during decision-making indicates that positive moods often enhance creativity and problem-solving abilities, while negative moods can improve analytical and evaluative thinking~\cite{schwartz2002maximizing}. 
However, prolonged negative emotional states, such as stress, can impair decision quality, particularly in high-pressure situations~\cite{weik2009collapse}. 
These insights are crucial for understanding how individual emotional well-being influences the collaborative production of maps.

People aside, other factors such as organizational protocols and guidelines may also shape decision-making processes. Stricter organizational practices, such as regulatory standards like GDPR~\cite{gdpr2016} or HIPAA~\cite{hipaa1996}, or detailed internal procedures for data management, often impose rigid decision hierarchies that reduce the likelihood of irregularities but may stifle creativity and innovation~\cite{amabile1996assessing}.
Conversely, more lenient practices, such as those found in innovation labs or start-ups, can encourage creativity and risk-taking but may lead to inconsistencies in outputs~\cite{march1993organizations}.
With respect to mapmaking, organizations with well-defined governance structures for map production may better maintain data accuracy and ethical standards, while looser governance might lead to exploratory but error-prone outputs.

In this work, we interview mapmakers to understand how they utilize their training and preferences, and also organizational structures and norms (protocols and guidelines), to make decisions related to mapmaking.

%% file: src/sections/03-expert-interviews.tex
We used a qualitative interview study methodology~\cite{weiss1995learning}, drawing on the protocols, questions, and analysis techniques from prior studies~\cite{kandel2012enterprise, muller2019data, madanagopal2019analytic} in visualization, CSCW, and HCI literature.

\subsection{Participants and Organizations}
We interviewed 16 cartographers and GIS experts (\expert{1-16}, Table~\ref{tab:expert-analysis-table}) with a self-reported mapmaking experience between 3--33 years (median: 15, mean: 15.43). Each of these 16 mapmakers worked with one of 13 (\org{1-13}) globally-serving government organizations and NGOs (similar to the World Health Organization (WHO)) and federal agencies (similar to the Department of Agriculture) based in the U.S. (eleven) and India (two). There were three mapmakers from the same organization, and two mapmakers from another organization\footnote{We cannot disclose exact details about the organizations, their employee mapmakers, and the maps they make as it can break anonymity.}.

These mapmakers were in the \emph{25--34}~(4), \emph{35--44}~(7), \emph{45--54}~(4), or \emph{55 and above}~(1) age groups (in years) and reported \emph{female}~(6) and \emph{male}~(10) genders. Our experts' highest educational qualification comprised \emph{bachelors}~(1), \emph{masters}~(13), or \emph{doctoral}~(2) degrees in \emph{geography}~(9), \emph{public health}~(3), \emph{geology}~(2), \emph{applied geology}~(1), \emph{geographic information systems and technology}~(1), \emph{sustainable development}~(1), \emph{civil engineering}~(1), and \emph{computer science}~(1); four mapmakers have dual degrees. 
Based on their job titles and their own task descriptions, we categorized these mapmakers into three primary roles: \emph{Creators} who created maps by themselves (n=10), \emph{Managers} who managed the creators (4), and \emph{Reviewers} who reviewed maps sent to them (2).

We selected these organizations using purposive sampling~\cite{patton2014qualitative} -- a non-probabilistic method where participants are intentionally chosen based on specific criteria. We stopped recruitment after interviewing individuals from 13 organizations as we achieved empirical saturation -- the point at which additional interviews were likely to elicit insights that had already been noted~\cite{guest2006how, morse2000determining}. We determined saturation through concurrent analysis, ensuring that the perspectives collected sufficiently addressed our research questions and objectives.

\subsection{Recruitment and Interviews}
We contacted 15 organizations primarily through their \emph{contact us} form(s) and email address(es) on their website. 
One organization declined and another did not respond.
The ones that did respond, often re-routed us, sometimes multiple times, across departments or even organizations before eventually putting in touch with the appropriate mapmakers.
Out of the 20 mapmakers who responded to these calls, four declined our interview request.
Thus, we eventually interviewed 16 mapmakers from 13 organizations.
All interviews (except one telephone interview) were conducted via Zoom, with either two or three study administrators in attendance.
All interviews lasted about forty-five minutes and were screen- and audio-recorded. 
There was no compensation for participation and consent was provided verbally. 
We asked our interviewees a set of semi-structured questions (reproduced below) and asked them to fill out a short Qualtrics~\cite{qualtrics} demographic questionnaire\footnote{Age Group, Gender, Educational Qualification, Job Title, and Mapmaking Experience. This demographic questionnaire is in supplemental material.}. The purpose of questions from (2), where the interviewee describes a map they made, is to capture how the mapmaker explains the process of a map where we can see the outcome. 

\begin{enumerate}[nosep]
    \item \emph{How often do you create choropleth maps?}
    \item \emph{What is a notable or recent choropleth map you have made?}
        \begin{enumerate}[nosep]
            \item \emph{What was it about? (data)}
            \item \emph{Who was it for? (client)}
            \item \emph{Who was the audience? (end-user)}
            \item \emph{How was the map communicated? (media, format)}
        \end{enumerate}
    \item \emph{Can you tell us more about your mapmaking process?}
        \begin{enumerate}[nosep]
            \item \emph{Which tools do you typically use?}
            \item \emph{Do you look at the data before making the map? How?}
            \item \emph{How do you determine the bins (or classes)?}
            \item \emph{What kinds of binning methods do you like to use?}
            \item \emph{How do you decide the colors?}
            \item \emph{Are there rules of thumb that you use? }
        \end{enumerate}
    \item \emph{Do you tend to use the same mapmaking process every time? If not, why and what's different?}
    \item \emph{Does your organization have rules or protocols, e.g., which binning methods or color schemes to (not) use?}
\end{enumerate}

\subsection{Analysis}
We transcribed the audio recordings, divided transcripts into smaller sections, and applied open coding~\cite{boyatzis1998transforming}.
When coding a section, we applied constant comparison~\cite{strauss1998basics} by reviewing the data to compare it with other sections and then labeled similar sections into the same category, to group related categories into a theme. We used theoretical sampling~\cite{strauss1998basics} by considering categories that had not yet emerged and refined these through discussions among the study team.

%% file: src/sections/04-expert-interviews-findings.tex
In this section, we describe how mapmakers' cartographic training and preferences interplay with organizational structures while making (choropleth) maps. We first provide an overview of the mapmakers' descriptions of notable/recent maps they have made at these organizations along with the map's audience, purpose, format, and type.
Next, we discuss the organizations' mapping workflows and the tools and libraries used.
Following, we discuss their choropleth mapmaking process describing if and how they prepared the data for analysis and determined binning methods and colors.
Finally, we discuss the impact of organizational structures and norms such as protocols and guidelines on mapmaking. In each of the following subsections, we describe both individual and collaborative efforts of mapmaking focusing on the cartographer as a trained practitioner working with others in and as part of an organizational ecosystem.

\subsection{Notable/Recent Maps, Audience, Purpose, Formats, and Types} 
Our interviewees' stated notable/recent maps were about a variety of topics: housing status, groundwater level, vaccine hesitancy, electricity distribution, lake bathymetry, economic factors, medicaid recipients, gender pay gap, energy consumption, climate and environmental justice, wildlife population, poverty and nighttime lights, archaeological stupas (structures), population literacy, and population density.
While the purpose of our question on recent/notable maps was to serve as a sanity check and start the conversation, our interviewees went on to detail all types of maps they typically make.
Our interviewees' goals were to (1) show absolute values of one (univariate) or two (bivariate) non-spatial attribute values on a map, (2) facilitate comparisons across geographic regions by coloring them relative to one or more user-defined thresholds, or (3) demarcate neighboring geographic regions by coloring them distinctly (e.g., ``Zone 1'' or ``Green Zone''). 
These maps were used to elicit questions, aid policy making, facilitate teaching, promote research, and even serve ornamental purposes.
These maps served general audiences as well as teachers, students, researchers, tourists, village panchayats, forest rangers, and government officials. All organizations published these maps in print and online media, in static (e.g., PDF, raster) and/or interactive (e.g., SVG, vector) formats, for usage across smartphones, desktops, paper, and/or braille displays.

\subsection{Mapping Workflows} 
\begin{figure}[!ht]{
    \centering
    \includegraphics[width=\textwidth]{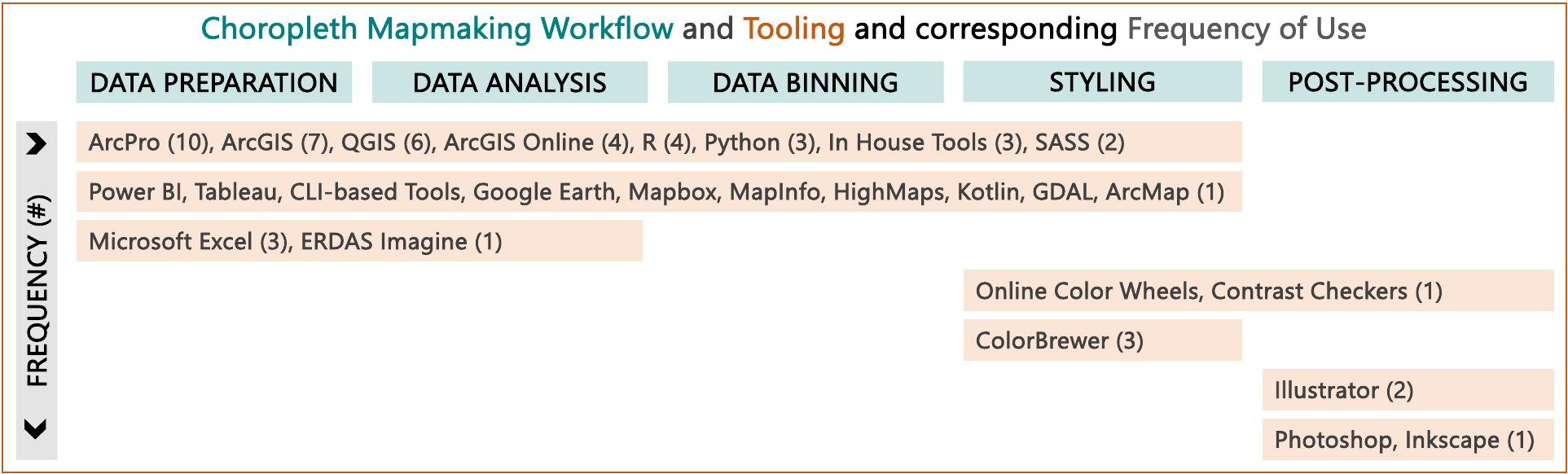}
    \caption{Choropleth mapmaking workflow and associated tooling and corresponding usage frequency.}
    \Description{Shows a scatterplot-like chart showing the five stages of a choropleth mapmaking workflow on the X axis: Data Preparation, Data Analysis, Data Binning, Styling, Post-Processing; and the frequency of tools used by mapmakers on the Y axis. The different tools (e.g., ArcPro, Python, R) are accordingly positioned in the main chart area depending on the workflow stage they are used and the number of mapmakers who mentioned using it. For example, ArcPro is most popular, used from Data Preparation through Styling; Illustrator and Photoshop are popular Post-Processing tools.}
    \label{fig:tools}
  }
\end{figure}

From our interviewees' descriptions, we derived a general, high-level mapmaking workflow comprising five steps
along with the tools used during each step (Figure~\ref{fig:tools} and Figure~\ref{fig:binFrequency}b).

\begin{enumerate}[nosep]
    \item \emph{Data preparation} (or pre-processing) involves collection, cleaning, transforming, and formatting raw data into a format suitable for analysis~\cite{zhang2003data}.
    \item \emph{Data analysis} involves inspecting the data distribution (e.g., to find outliers) to inform next steps (e.g., binning).
    \item \emph{Data binning} involves grouping data values into few, well-defined bins before visualizing on a map as they make it easier to highlight data patterns and trends across geographic regions. 
    \item \emph{Map styling} involves determination of design choices to effectively communicate information, e.g., selecting a color scheme and the use of additional visual aids such as legends and labels.
    \item \emph{Map post-processing} involves enhancing the map's visual representation by adjusting colors, labels, and other elements to improve readability.
\end{enumerate}

Workflows often varied across individuals and organizations and may include additional steps (e.g., \emph{reviews}), discard certain steps (e.g., no \emph{post-processing}), follow a different sequence of steps (e.g., \emph{styling} before \emph{binning}), and/or involve iterations (e.g., re-\emph{analysis} after \emph{binning}) after discussions with colleagues, clients, and vendors. 
For example, \expert{6}'s \emph{``process deviated a lot as there was no GIS department and folks have different skills so [the organization only tries] to encourage same colors, legend styles, and product brand styles.''}
Along similar lines, \expert{16}'s organization exercises no oversight, affording complete autonomy for mapmaking. \org{2}, however, has \emph{``pretty well-defined rules for all kinds of graphics displays ranging from standards and procedures.''}
Notably, no mapmaker mentioned revising a map \emph{after} publication, based on feedback from the target audience (e.g., through the published map interface).

\subsection{Mapping Tools and Libraries}
Our interviewee mapmakers often used a combination of tools depending on various factors including the size and type of the organization, size and nature of the underlying data, team dynamics in the work environment, intended use-case(s) and target audience, tool characteristics (e.g., pricing matrix, licensing terms, technical specifications), personal expertise, and experience working with the tool(s).
Table~\ref{tab:expert-analysis-table}, Figure~\ref{fig:tools}, Figure~\ref{fig:binFrequency}b show the tools mapmakers reported using for mapmaking.
Proprietary ESRI~\cite{esri} products: ArcGIS, ArcPro, ArcGIS Online were the most popular followed by their open-source alternative, QGIS; the choice between the two was a result of the organization's size, available resources, and nature of work.
\expert{3, 13} used in-house mapping tools whereas \expert{15} used a customized version of ArcMap.
Scripting with R, Python, and Command Line Interface-based tools was also popular to automate the creation of interactive maps.
Other commonly used tools included Excel (for data preparation and analysis); ColorBrewer, web-based color wheels, and in-house color palettes for determining color schemes; and Illustrator, Photoshop, and Inkscape for blending, after effects, and other post-processing.

\subsection{Data Preparation \& Analysis: Looking at Data before Mapping}
Twelve mapmakers said they always looked at the data either in its raw state or as a visualization before mapping as subsequent decisions \emph{``always depended on the data''} (\expert{9-11}).
\expert{5} specifically checked for outliers or skewed distributions to determine appropriate binning methods to apply next.
\expert{4} checked for a mid-point around which the data may be diverging.
\expert{12} reviewed the data \emph{``to try and tell an appropriate story.''} \expert{12} also noted that \emph{``data and time gaps [e.g., missing values] may exist so [assessing for data quality] helps determine if the data is complete enough [to be mapped]. Only then do [they] bring data into the GIS software to work it out.
''}
\expert{16} looked at the data without performing robust statistical analysis, adding, \emph{``shame on me, I should.''}

Of the 12 mapmakers who `always' looked at the data, six leveraged the histogram feature in ArcPro (\expert{1,2,9-11,16}) or QGIS (\expert{4}), five used Excel's pivot table, sorting, and/or filtering affordances (\expert{2,5,7,12,16}), and one mapmaker each wrote R (\expert{1}), SASS (\expert{1}), and Python (\expert{5}) scripts.
\expert{2} also studied qualitative aspects of data, e.g., its source and how it was collected.
The remaining four mapmakers either looked at data sometimes or never.
Of these, \expert{6,8,15} often directly started making maps without looking at the tabular data beforehand, as \emph{``that is much more fun''} (\expert{8}).
\expert{14} never look at data but remarked, \emph{``you're making me think I should.''}

\subsection{Determining Data Binning Methods}

Our mapmakers determined bins in many viable ways with some choices more popular than others, as described next.

\bpstart{Criteria.} Our mapmakers chose a binning method based on a number of factors including the underlying data distribution (e.g., use \emph{standard deviation} for normal data distributions--\expert{1}),
organizational guidelines (e.g., always use the \emph{standard deviation} method--\expert{1}), target audience's literacy (e.g., use \emph{pretty breaks} to present easy-to-understand rounded bin extents--\expert{4-8,10,12,14,15}), prior work (e.g., use the same bins for recurring maps--\expert{6,7}), map's narrative (use \emph{manual interval}--\expert{7,10,11,14}), or personal preference (e.g., use the ArcGIS default of \emph{natural breaks} with five bins--\expert{16}).

\paragraphHeadingSpace\bpstart{Popular binning methods.} 
Among binning methods, our mapmakers used \emph{pretty breaks} and \emph{manual interval} most often, followed by \emph{natural breaks}, \emph{quantile}, \emph{standard deviation}, \emph{unclassed}, and \emph{equal interval}, similar to Brewer~et~al.'s evaluation~\cite{brewer2002evaluation} (Figure~\ref{fig:binFrequency}c, Table ~\ref{tab:binning-methods-table}). Figure~\ref{fig:binningmethods-figure} shows the outputs of these binning methods. We believe these methods are still very popular as they are an integral part of cartographic education, are readily available in popular tools, and also relatively easier to implement.

The overall variance was due to several factors at play such as personal preferences (\emph{``I just use the ArcGIS default, natural breaks''}--\expert{16}), 
organizational policies (\emph{``it is a practice to use standard deviation here''}--\expert{1}), 
prior choices (\emph{``if there is a report that goes out every year, we also re-use the binning method every year, irrespective of the new data distribution''}--\expert{6}), 
and on the data (\emph{``equal interval is used for soft-range data, geometric interval or manual interval is used for long-range data''}--\expert{3}).

\input{src/tables/expert-analysis.tex}

\input{src/tables/binning-methods.tex}

\paragraphHeadingSpace\bpstart{How many bins.} Table~\ref{tab:expert-analysis-table} and Figure~\ref{fig:binFrequency}a also show the distribution of the number of bins most commonly used by mapmakers for univariate data. These mapmakers generally created between 3 and 11 bins with 5 being the most common choice; one mapmaker recalled having used 18 bins once for a land-use map; three mapmakers mentioned using `any' number of bins, indicating flexibility and dependence on the circumstance. 
\expert{14} said they prefer \emph{``bins that make intuitive sense so [they] don't like to use the automatic natural breaks or the quantile methods that are designed to make the map look pretty.''}
This finding mostly aligns with previous recommendations and practices to use 5--7 bins~\cite{mersey1990colour, declerq1995choropleth, axismaps, brewer2002evaluation}.

\paragraphHeadingSpace\bpstart{Manually determining bins is common.} Mapmakers often manually specify bins to better cater to the map's underlying data, intended purpose, and target audience.
\expert{10} mentioned that \emph{``sometimes, if there is a bin $<250$ but there is only one district in that bin [e.g., outlier], then we merge it with the next bin to ensure a better, more even distribution.''} 
\expert{13}, similarly, \emph{``once ran into an issue with some outlier data points [that were] throwing off the scale. [They] had to change some [bin extents] to be greater than 80\%, because it just had like two or so counties. [They] iterated with the research team on the best practices on how to achieve this washing of one end without skewing the data.''} 
\expert{6,15} followed a hybrid approach by first choosing one of the \emph{natural beaks} (\expert{15}) or \emph{standard deviation} (\expert{6,15}) binning methods, then tweaking the bins by themselves (\expert{15}) or with the help of subject-matter experts (\expert{6}) to achieve the best compromise between statistical accuracy and visual saliency and acuity.

\paragraphHeadingSpace\bpstart{Many unpredictable factors exist} that can impact binning-related decision-making.
For example, \expert{5} said their process often depends on their mood and how they're feeling on the day, e.g., \emph{``some days [they will] want to be super logical and organized, document everything, try and back-up all choices; and some other days [they] just want to get this map `out of the door'.''}
Next, \expert{15} said that one of their clients once reviewed a map draft and asked them to \emph{``change [certain bin extents] so that [certain] districts show as bright red, so that people think there's the problem, which defeats the purpose of making an unbiased map.''} 
\expert{15} also mentioned sometimes having to \emph{```fight with [the client]', but most of the time the client [gave in to the mapmaker's expertise].''}

\paragraphHeadingSpace\bpstart{Mapmakers do not always bin their data.} 
\expert{4} used the \emph{unclassed} method (dividing data into a continuous range of values, rather than into discrete bins) unless they have deep knowledge about the data only \emph{``to present data as-is without [infusing] additional biases that can influence the readers.''}
\expert{5} also used the \emph{unclassed} method for data with errors because \emph{``if you [bin] based on point estimates, then [binning] may amplify the errors.''}

\paragraphHeadingSpace\bpstart{Rules of thumb.} 
Most mapmakers mentioned not having too many bins, or having rounded, even, or familiar bin extents (\expert{4-8,10,12,14,15}) which are easier to interpret.
\expert{15} generally opts for 4--5 bins but there \emph{``might not be space in the legend to show all bins, so that's also a consideration.''}
\expert{3} noted making decisions based on literacy of the target audience, \emph{``3--4 bins are good for a non-technical audience whereas subject-matter experts may need bigger ranges; [in fact,] land use data sometimes have 18 bins.''}
\expert{6,14,15} are comfortable defining more bins for a divergent (not sequential) color scale.
These align with existing literature discussed in Section~\ref{sec:principles}.
\expert{1}'s rule of thumb was to use the `Albers USA' conic map projection for static U.S. maps and Web Mercator for other geographies.

\subsection{Determining Colors}

\bpstart{Criteria.} Mapmakers determined colors based on various criteria including \emph{personal preferences} (\expert{10,16}), \emph{collaborative brainstorming with their teams} (\expert{1}), \emph{data-driven considerations} around its semantics (\expert{11}) and type (\expert{5}), \emph{target audience literacies} (\expert{3,8,12}), \emph{socio-economic connotations and political implications} (\expert{1,13}), \emph{accessibility requirements} in terms of using colorblind (\expert{1,2,4-7,12-16}) and print (\expert{3,5,16}) friendly colors, \emph{prior similar work} (\expert{1,7}), \emph{client requirements} (\expert{15}), and \emph{branding requirements} (\expert{1,4-6,12,16}). 

\paragraphHeadingSpace\bpstart{Tools and Strategies.} 
Mapmakers often utilized existing color palettes as-is or customized them using existing GIS tools such as ArcGIS (\expert{10,16}), custom online color wheels (\expert{2}) and palettes such as ColorBrewer~\cite{harrower2003colorbrewer} (\expert{4,5}), and organizational stylesheets (\expert{4,6,16}), including personal collections of accessible color palettes (\expert{4}).
\expert{10} \emph{``chose color schemes that looked soothing to [their] eyes''} and would often play around with color tones and saturation, and/or mix colors, e.g., \emph{``50\% magenta, 10\% cyan and so on.''}
For \expert{10}, \emph{``determining colors depends only on [them], not [their] boss''}
whereas for \expert{11}, \emph{``[it] always depends on the data and superiors.''}
For \expert{1}, it is \emph{``an iterative, cooperative process with their mapping team, an internal communications team, and an additional external media firm that helps with branding, e.g., to colors that are linked to those on the website.''}
\expert{1,6,7} also replicated the color schemes from relevant prior projects for consistency.

\paragraphHeadingSpace\bpstart{Semantics matter.} Mapmakers often associated intuitive colors with relatable entities, e.g., forest is green, land is brown, water is blue. 
Mapmakers also avoided colors with negative connotations (which can change the effect of a map \cite{anderson2021affective}), e.g., \expert{1} never uses red to connote racial groups and \expert{13} avoid reds and blues due to their political implications in the U.S. 
\expert{14} often chose pink \emph{``for binary data as it doesn't have many semantic associations.''}

\paragraphHeadingSpace\bpstart{Conflicts arise.} Sometimes, however, these considerations conflicted with one another, resulting in violations or requiring workarounds and compromises. For instance, \expert{6} \emph{``couldn't always show natural gas in blue and ground in orange as it depends on what other data was being shown.''}
\expert{7} was \emph{``often unable to choose colorblind-friendly colors and used hatching as a workaround.''}

\subsection{Organizational Structure and Team Dynamics}
Our interviewee organizations made maps for either their own target audience or external clients based on the latter's requirements.
To achieve this, mapmakers often worked independently or iterated with their own (\expert{1,11}) or other teams in the organization (\expert{1,6,13}), including external clients (\expert{15}) and vendor firms (\expert{1}).
For instance, \expert{1} had iterative, collaborative brainstorming sessions with colleagues to determine colors and binning methods.
\expert{11} either followed instructions from their superiors or pursued own opinions before verifying with them.
\expert{13}'s team received an initial binning method, color scheme, and a map story from a research team and then iterated with them.
\expert{6} similarly consulted subject-matter experts but sometimes determined bins by themselves.
\expert{1} also consulted an internal communications team and an external media firm to style maps while keeping branding-related considerations in mind.
\expert{15} worked with an editing team that designed the final report and the map.
\org{2,5,6,8-12} had review teams to ensure compliance with protocols.

\subsection{Influence of Organizational Protocols and Guidelines}
All organizations except \org{3} reported having some kinds of mapmaking protocols, guidelines, and/or policies that are enforced for all public facing maps used for presentation, publication, or sale.
These may be desirable from a consistency, credibility, and/or brand promotion standpoint.

\paragraphHeadingSpace\bpstart{Types of protocols and guidelines.} 
\org{2,4,6,7,8,11,13} had well-defined rules and style guides on colors, symbology, typology, and/or other essential map elements such as the date, credit(s), disclaimer(s), data source(s), and/or legend.
\org{1,5} followed loose guidelines around map projections and data binning methods and were currently in the process of establishing more formal cartographic guidelines.
\expert{14} mentioned having to comply with U.S. Section 508~\cite{section508} on accessibility.
\org{2,5,6,8-12} underwent reviews before presenting, publishing, and/or selling maps. 
For example, \org{9,10,12} must ensure global (disputed) administrative boundaries are in compliance with international laws. 
Hence, \org{10} only uses shapefiles created and maintained by their in-house GIS team. 
Notably, \expert{15}'s organization is not permitted to make maps for certain countries, e.g., India.

\paragraphHeadingSpace\bpstart{Strict or lenient?} \expert{8}'s organization had guidelines on colors and provisions for creative freedom; \expert{8} tried \emph{``to mix things up a little bit, e.g., by not using the same red or gray for roads or the same shade of blue for water.''}
\expert{6}'s organization was stricter and they \emph{``must explain [to the review team] why [they] tweaked colors. It [was] kind of annoying to be limited but it makes sense in the long run.''}
\expert{16} was not content with the fonts mandated by their organization as, \emph{``some of [them] are not readily available and require a paid subscription. 90\% of [their] staff doesn't have it so it's kind of useless.''}
\expert{4} were more wishful, \emph{``There are no style guides and that's something I'm annoyed about.''}

%% file: src/tables/expert-analysis.tex
\begin{table}[t]
    \footnotesize
    \setlength{\tabcolsep}{0pt}
    \centering
    \begin{tabular}{ccclcl}
        \hline
        \multirow{2}{*}{\begin{tabular}[c]{c}\textbf{Mapmaker}\\\small{(1--16)}\end{tabular}} &
        \multirow{2}{*}{\begin{tabular}[c]{c}\textbf{Experience}\\\small{(3--33 years)}\end{tabular}} &
        \multirow{2}{*}{\begin{tabular}[c]{c}\textbf{Role}\\\small{(three)}\end{tabular}} &
        \multirow{2}{*}{\begin{tabular}[c]{l}\textbf{Tools}\\\small{(multiple)}\end{tabular}} &
        \multirow{2}{*}{\begin{tabular}[c]{c}\textbf{\# Bins}\\\small{(custom)}\end{tabular}} &
        \multirow{2}{*}{\begin{tabular}[c]{l}\textbf{Binning Methods}\\\small{(twelve)}\end{tabular}} \\ \\
        \hline

        \multirow{1}{*}{\expert{10}} & 
        \multirow{1}{*}{33} & 
        \multirow{1}{*}{Creator} &
        \multirow{1}{*}{
            \begin{tabular}[c]{l}
                ArcGIS, ArcPro
            \end{tabular}
        } &
        \multirow{1}{*}{5--10} &
        \multirow{2}{*}{
            \begin{tabular}[c]{l}
                \{manual, equal\} interval,\\ natural breaks, pretty breaks
            \end{tabular}    
        }  \\ \\

        \multirow{1}{*}{\expert{3}} & 
        \multirow{1}{*}{18} & 
        \multirow{1}{*}{Creator} &
        \multirow{2}{*}{
            \begin{tabular}[c]{l}
                ArcPro, QGIS, Kotlin,\\ In-house tool
            \end{tabular}
        } &
        \multirow{1}{*}{3--6, 18} &
        \multirow{3}{*}{
            \begin{tabular}[c]{l}
                \{equal, manual, geometric\} \\interval, percentile, ck-means,\\ natural breaks, boxplot
            \end{tabular}
        }  \\ \\ \\

        \multirow{1}{*}{\expert{4}} & 
        \multirow{1}{*}{15} & 
        \multirow{1}{*}{Creator} &
        \multirow{2}{*}{
            \begin{tabular}[c]{l}
                QGIS, Python, GDAL~\cite{gdal},\\ CLI-based tools, 
            \end{tabular}
        } &
        \multirow{1}{*}{3--9} &
        \multirow{2}{*}{
            \begin{tabular}[c]{l}
                unclassed, pretty breaks,\\ equal interval
            \end{tabular}    
        }  \\ \\

        \multirow{1}{*}{\expert{5}} & 
        \multirow{1}{*}{15} & 
        \multirow{1}{*}{Creator} &
        \multirow{3}{*}{
            \begin{tabular}[c]{l}
                ArcPro, Python, Excel, R,\\ ColorBrewer, Inkscape~\cite{inkscape},\\ Online contrast checkers
            \end{tabular}
        } &
        \multirow{1}{*}{5--7} &
        \multirow{1}{*}{
            \begin{tabular}[c]{l}
                unclassed, manual interval,\\ pretty breaks
            \end{tabular}
        }  \\ \\ \\

        \multirow{1}{*}{\expert{15}} & 
        \multirow{1}{*}{15} & 
        \multirow{1}{*}{Creator} &
        \multirow{2}{*}{
            \begin{tabular}[c]{l}
                ArcGIS, ArcGIS Online,\\ Python, QGIS, Illustrator
            \end{tabular}
        } &
        \multirow{1}{*}{4, 5} &
        \multirow{2}{*}{
            \begin{tabular}[c]{l}
                standard deviation, pretty breaks,\\ natural breaks, manual interval
            \end{tabular}
        }  \\ \\

        \multirow{1}{*}{\expert{16}} & 
        \multirow{1}{*}{10} & 
        \multirow{1}{*}{Creator} &
        \multirow{1}{*}{
            \begin{tabular}[c]{l}
                ArcPro, ArcGIS Online, Excel
            \end{tabular}
        } &
        \multirow{1}{*}{5} &
        \multirow{1}{*}{ natural breaks}  \\

        \multirow{1}{*}{\expert{11}} & 
        \multirow{1}{*}{7} & 
        \multirow{1}{*}{Creator} &
        \multirow{1}{*}{
            \begin{tabular}[c]{l}
                ArcGIS
            \end{tabular}
        } &
        \multirow{1}{*}{5--7} &
        \multirow{1}{*}{
            \begin{tabular}[c]{l}
                unclassed, manual interval
            \end{tabular}    
        }  \\

        \multirow{1}{*}{\expert{8}} & 
        \multirow{1}{*}{6} & 
        \multirow{1}{*}{Creator} &
        \multirow{1}{*}{
            \begin{tabular}[c]{l}
                ArcGIS, ArcMap
            \end{tabular}
        } &
        \multirow{1}{*}{any} &
        \multirow{1}{*}{
            \begin{tabular}[c]{l}
                defined interval, pretty breaks
            \end{tabular}
        }  \\

        \multirow{1}{*}{\expert{1}} & 
        \multirow{1}{*}{4} & 
        \multirow{1}{*}{Creator} &
        \multirow{2}{*}{
            \begin{tabular}[c]{l}
                ArcPro, ArcGIS Online, R,\\ SASS, ColorBrewer
            \end{tabular}
        } &
        \multirow{1}{*}{5--6} &
        \multirow{1}{*}{
            \begin{tabular}[c]{l}
                standard deviation, quantile
            \end{tabular}    
        } \\ \\

        \multirow{1}{*}{\expert{14}} & 
        \multirow{1}{*}{3} & 
        \multirow{1}{*}{Creator} &
        \multirow{1}{*}{
            \begin{tabular}[c]{l}
                ArcPro, R
            \end{tabular}
        } &
        \multirow{1}{*}{4, 5} &
        \multirow{2}{*}{
            \begin{tabular}[c]{l}
                manual interval, quantile,\\ pretty breaks
            \end{tabular}
        }  \\ \\[0.5ex]

        \hline

        \multirow{1}{*}{\expert{6}} & 
        \multirow{1}{*}{30} & 
        \multirow{1}{*}{Manager} &
        \multirow{3}{*}{
            \begin{tabular}[c]{l}
                ArcPro, ArcGIS Online, QGIS,\\ 
                SASS, R, HighMaps~\cite{highmaps},\\ PowerBI~\cite{powerbi}, Tableau~\cite{tableau}
            \end{tabular}
        } &
        \multirow{1}{*}{5--7} &
        \multirow{2}{*}{
            \begin{tabular}[c]{l}
                manual interval, pretty breaks, \\standard deviation, quantile
            \end{tabular}    
        }  \\ \\ \\

        \multirow{1}{*}{\expert{2}} & 
        \multirow{1}{*}{25} & 
        \multirow{1}{*}{Manager} &
        \multirow{3}{*}{
            \begin{tabular}[c]{l}
                ArcGIS, Excel, Photoshop~\cite{photoshop},\\ Illustrator~\cite{illustrator}, MapInfo~\cite{mapinfo},\\ Online color wheels
            \end{tabular}
        } &
        \multirow{1}{*}{3--5} &
        \multirow{1}{*}{
            \begin{tabular}[c]{l}
                natural breaks, quantile
            \end{tabular}
        } \\ \\ \\

        \multirow{1}{*}{\expert{9}} & 
        \multirow{1}{*}{25} & 
        \multirow{1}{*}{Manager} &
        \multirow{1}{*}{
            \begin{tabular}[c]{l}
                ArcGIS, ArcPro, ERDAS Imagine~\cite{erdas}
            \end{tabular}
        } &
        \multirow{1}{*}{any} &
        \multirow{1}{*}{
            \begin{tabular}[c]{l}
                \emph{``I let my team decide this.''}
            \end{tabular}
        }  \\

        \multirow{1}{*}{\expert{13}} & 
        \multirow{1}{*}{6} & 
        \multirow{1}{*}{Manager} &
        \multirow{1}{*}{
            \begin{tabular}[c]{l}
                In-house tool, MapBox, ColorBrewer
            \end{tabular}
        } &
        \multirow{1}{*}{9, 11} &
        \multirow{1}{*}{
            \begin{tabular}[c]{l}
                manual interval
            \end{tabular}
        }  \\ [0.5ex]

        \hline 

        \multirow{1}{*}{\expert{7}} & 
        \multirow{1}{*}{20} & 
        \multirow{1}{*}{Reviewer} &
        \multirow{2}{*}{
            \begin{tabular}[c]{l}
                QGIS, Google Earth~\cite{googleearth},\\ ArcPro
            \end{tabular}
        } &
        \multirow{1}{*}{any} &
        \multirow{2}{*}{
            \begin{tabular}[c]{l}
                \{manual, equal\} \\interval, pretty breaks
            \end{tabular}
        }  \\ \\

        \multirow{1}{*}{\expert{12}} & 
        \multirow{1}{*}{15} & 
        \multirow{1}{*}{Reviewer} &
        \multirow{2}{*}{
            \begin{tabular}[c]{l}
                ArcGIS, ArcPro, QGIS,\\ Illustrator
            \end{tabular}
        } &
        \multirow{1}{*}{4--6} &
        \multirow{2}{*}{
            \begin{tabular}[c]{l}
                standard deviation, quantile,\\ pretty breaks
            \end{tabular}
        }  \\ \\[0.5ex] 
        
        \hline
    \end{tabular}%
    \caption{Summary statistics about our interviewee \textbf{Mapmakers'} \textbf{Experience} (in number of years), their \textbf{Role} in their respective organizations, and preferred \textbf{Tools}, \textbf{\# (number of) Bins}, and \textbf{Binning Methods} for making choropleth maps.}
    \label{tab:expert-analysis-table}

\end{table}

\begin{figure}[t]
    \includegraphics[width=\textwidth]{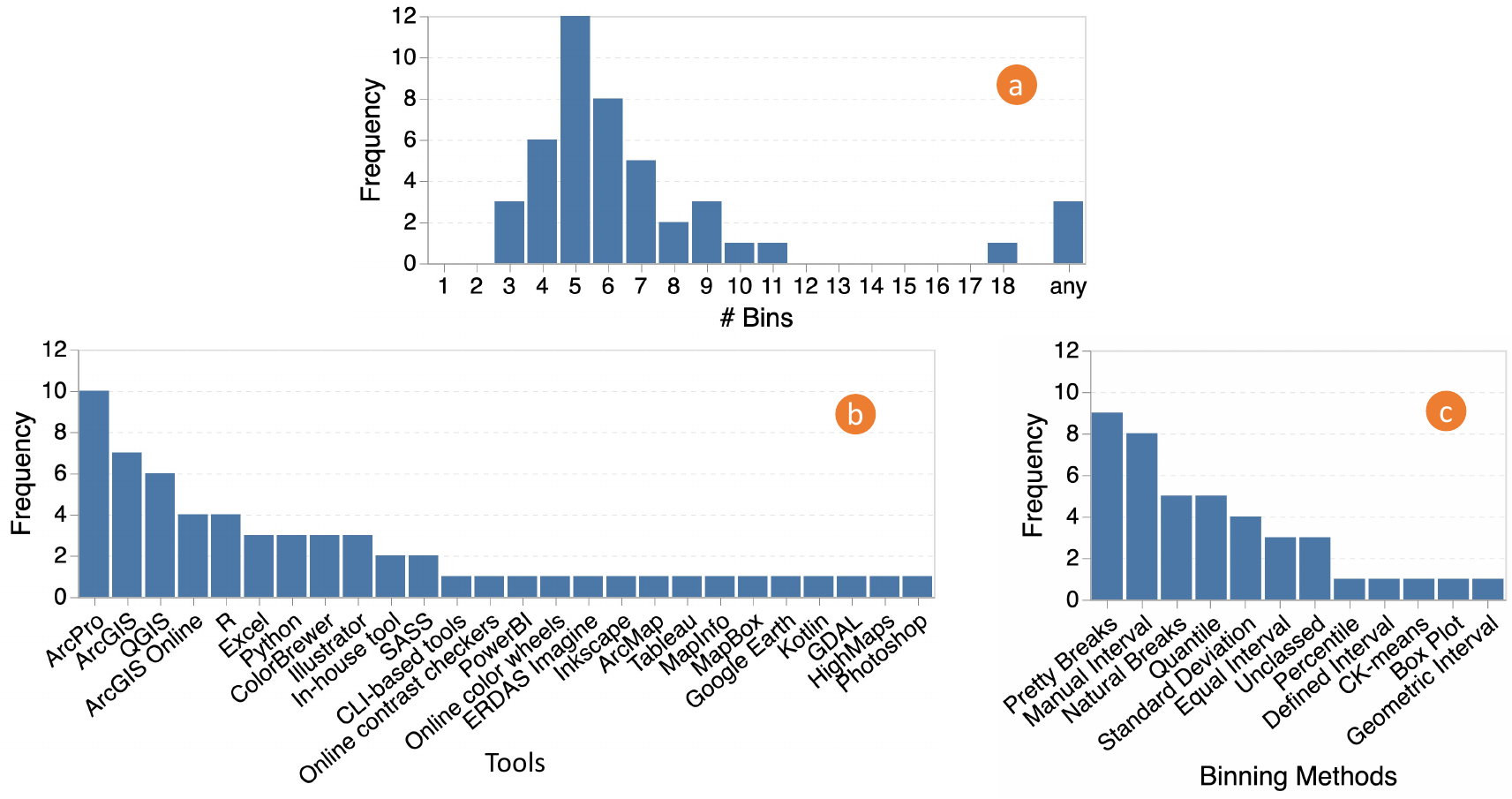}
    \caption{\textbf{Frequency} of (a) \textbf{\# (number of) Bins}, (b) \textbf{Tools}, and (c) \textbf{Binning Methods} most commonly used by our interviewee mapmakers, respectively; if a mapmaker mentioned they used `4--6' bins, we incremented the frequency for \textbf{\# Bins} = `4', `5', as well as `6'; if a mapmaker mentioned `any' (number of) bins, we represent this separately.}
    \Description{This figure shows three bar charts. The first bar chart shows the frequency of the number of bins most commonly used by mapmakers from 0 to 12 on the Y axis against the number of bins from 0 to 18 and a separate category titled "any" (number of bins) on the X axis. The bar chart peaks at X(number of bins) = 5 with a Y(frequency) = 12, followed by (X=6, Y=8) and (X=4, Y=6). Notably, none of our mapmakers used between 11 and 18 number of bins; only one once mapmaker used 18 bins; and three mapmakers used "any" number of bins. Note that if a mapmaker mentioned “4-6” bins we incremented the Y(frequency) for X(number of bins) = '4', '5', as well as '6'; the "any" response, as described earlier, is represented separately. The second bar chart shows the usage frequency of different tools from 0 to 12 on the Y axis against the different tools on the X axis. ArcPro (10), ArcGIS (7), QGIS (6) are the most common tools followed by ArcGIS Online (4) and R (4). Finally, the third bar chart shows the usage frequency of different binning methods from 0 to 12 on the Y axis against the different binning methods on the X axis. Pretty Breaks (9) and Manual Interval (8) are the most popular methods followed by Natural Breaks (5) and Quantile (5).}
    \label{fig:binFrequency}
\end{figure}

%% file: src/tables/binning-methods.tex
\begin{table}[t]
    \small
    \centering
    \begin{tabular}{ll}
        \hline
        \multirow{1}{*}{\begin{tabular}[c]{l}\textbf{Binning Method}\end{tabular}} &
        \multirow{1}{*}{\begin{tabular}[c]{l}\textbf{Definition}\end{tabular}} 
        \\
        \hline

        \multirow{1}{*}{Pretty Breaks~\cite{pysal}} & 
        \multirow{2}{*}{
            \begin{tabular}[c]{l}
                Rounds each bin break up or down into \emph{pretty} values that are easier to read,\\ understand, and explain, e.g., [0--10] instead of [0--9.364].
            \end{tabular}    
        } 
        \\ \\

        \multirow{1}{*}{Manual Interval~\cite{de2007geospatial}} & 
        \multirow{2}{*}{
            \begin{tabular}[c]{l}
                Organizes data into bins based on the analyst's predetermined criteria\\ or when automated methods are inappropriate or unavailable. 
            \end{tabular}    
        } 
        \\ \\

        \multirow{1}{*}{Natural Breaks~\cite{de2007geospatial}} & 
        \multirow{2}{*}{
            \begin{tabular}[c]{l}
                Minimizes the within-bin variance and maximizes the between-bin variance, \\ creating natural groupings of data that are more meaningful and interpretable.
            \end{tabular}    
        } 
        \\ \\

        \multirow{1}{*}{Quantile~\cite{de2007geospatial}} & 
        \multirow{1}{*}{
            \begin{tabular}[c]{l}
                Defines bin intervals such that their bin size is (approximately) the same.
            \end{tabular}    
        } 
        \\ [0.5ex]
        \hline
        
        \multirow{1}{*}{Standard Deviation~\cite{de2007geospatial}} & 
        \multirow{2}{*}{
            \begin{tabular}[c]{l}
                Defines bin extents at a certain number of standard deviations ($\sigma$) from the\\ mean ($\mu$) of the data, usually at intervals of 1.0$\sigma$ or 0.5$\sigma$.
            \end{tabular}    
        } 
        \\ \\

        \multirow{1}{*}{Equal Interval~\cite{de2007geospatial}} & 
        \multirow{1}{*}{
            \begin{tabular}[c]{l}
                Divides the data range into the specified bin count with equal bin intervals.
            \end{tabular}    
        } 
        \\
        
        \multirow{1}{*}{Unclassed~\cite{de2007geospatial}} & 
        \multirow{1}{*}{
            \begin{tabular}[c]{l}
                Divides data into a continuous range of values, rather than into discrete bins.
            \end{tabular}    
        }
        \\

        \multirow{1}{*}{Percentile~\cite{de2007geospatial}} & 
        \multirow{2}{*}{
            \begin{tabular}[c]{l}
                Defines bin intervals based on percentiles (\%):\\ e.g., [$0$,$1$), [$1$,$10$), [$10$,$50$), [$50$,$90$), [$90$,$99$), [$99$,$100$].
            \end{tabular}    
        } 
        \\ \\ 
        [0.5ex]
        \hline
        
        \multirow{1}{*}{Defined Interval~\cite{de2007geospatial}} & 
        \multirow{1}{*}{
            \begin{tabular}[c]{l}
                Divides the data range into bins based on a numeric bin interval.
            \end{tabular}    
        } 
        \\

        \multirow{1}{*}{CK-Means~\cite{wang2011ckmeans}} & 
        \multirow{1}{*}{
            \begin{tabular}[c]{l}
                Performs optimal one-dimensional binning using dynamic programming.
            \end{tabular}
        } 
        \\ 

        \multirow{1}{*}{Box Plot~\cite{tukey1977box, de2007geospatial}} & 
        \multirow{2}{*}{
            \begin{tabular}[c]{l}
                Typically defines six bins: four quartiles plus two classifications for data items\\ that are more than $1.5$ times the inter-quartile range (IQR) from the median.
            \end{tabular}    
        } 
        \\ \\
        
        \multirow{1}{*}{Geometric Interval~\cite{arcgis}} & 
        \multirow{2}{*}{
            \begin{tabular}[c]{l}
                Divides the data range into a specified bin count based on a geometric series\\ ($a$ + $ar$ + $ar^2$ + ...) where $a$ = constant and $r$ = geometric coefficient. 
            \end{tabular}    
        } 
        \\ \\ 
        [0.5ex] 
        \hline
        \\
    \end{tabular}%
    \caption{Catalog of \textbf{Binning Methods} (and their \textbf{Definitions}) used by our interviewee mapmakers.} 
    \label{tab:binning-methods-table}

    
\end{table}

\begin{figure}[t]
    \includegraphics[width=\columnwidth]{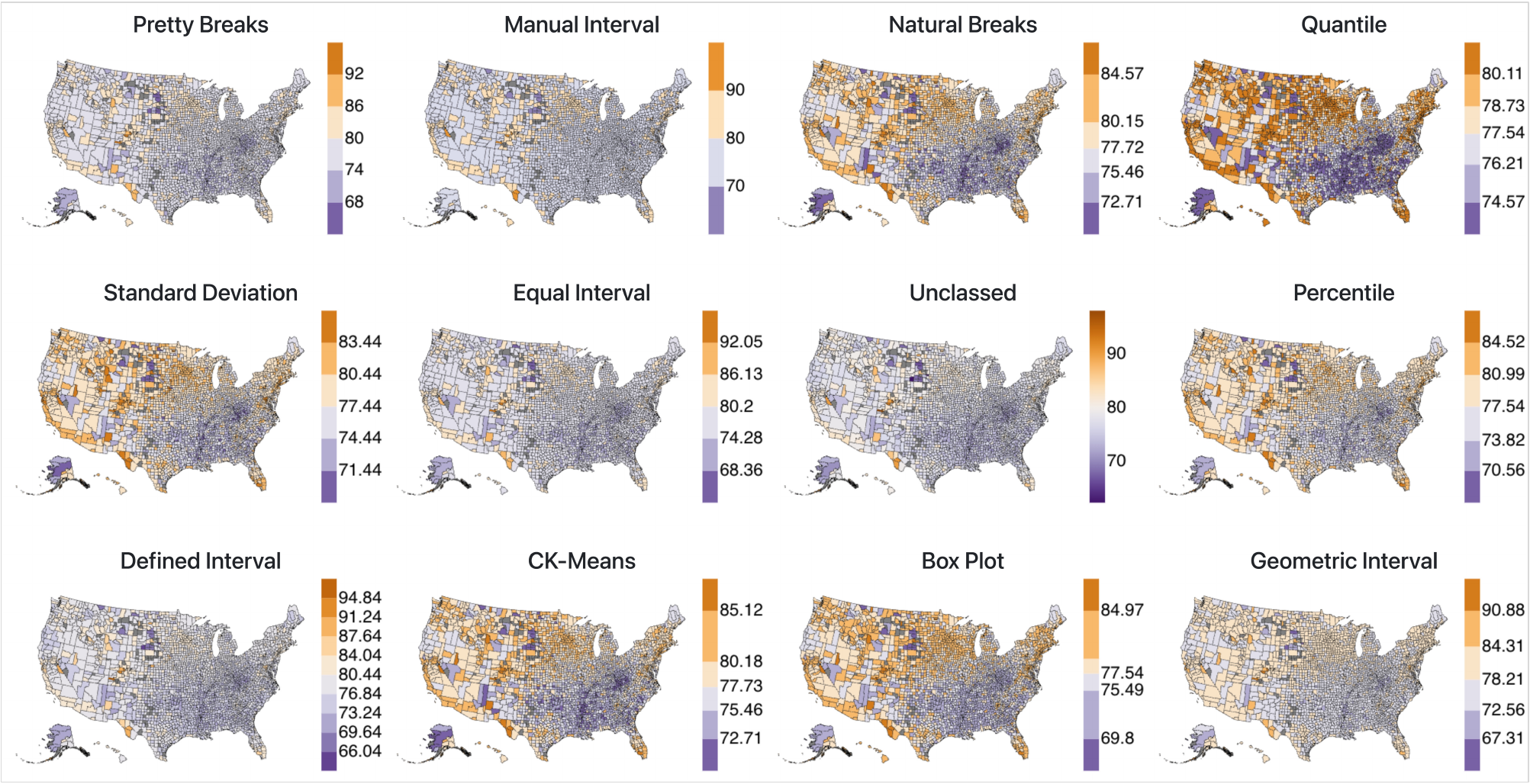}
    \caption{Small multiples of choropleth maps showing ``Life Expectancy'' (in years) among U.S. counties~\cite{cdc2022adultobesitychoropleth} across the twelve data binning methods (made using BinGuru~\cite{narechania2023resiliency} and Vega-Lite~\cite{satyanarayan2016vega}).} 
    \Description{Small multiples of choropleth maps showing the same data on 'Life Expectancy' (in years) among U.S. counties across the twelve data binning methods. The small multiples are positioned in a grid consisting of three rows with four columns each; first row shows Pretty Breaks, Manual Interval, Natural Breaks, and Quantile; second row shows Standard Deviation, Equal Interval, Unclassed, and Percentile; the third row shows Defined Interval, CK-Means, Box Plot, and Geometric Interval, in the decreasing order of their popularity (usage frequency). Each map is colored using a multi-hue divergent color scheme from Purples to Oranges.}
    \label{fig:binningmethods-figure}
  \end{figure}

%% file: src/sections/06-discussion.tex
\subsection{Enhancing Decision-Making in Mapmaking}
\paragraphHeadingSpace\bpstart{Navigating multiple choices.}
Evidenced in how mapmakers are telling us they make different decisions, there are many viable ways to make maps with some choices more popular than others.
For instance, mapmakers generally use 5--7 bins for univariate data, which mostly aligns with previous recommendations and practices~\cite{mersey1990colour, declerq1995choropleth, axismaps, brewer2002evaluation}. However, there may be other contextual factors such as the target audience, organizational norms, and mapmakers' `mood' and preferences that also influences associated decision-making.
While such takeaways can be useful for those seeking best practices for their data-driven products (e.g., choosing a useful number of data bins), future work could streamline this workflow and focus on developing guidelines and tools to help mapmakers navigate these choices and optimize their maps for specific purposes.
For example, developers could build a visualization recommendation system for a mapmaker to specify the to-be-mapped data, map purpose(s), capabilities of target audience, and other constraints; in response, the tool would suggest an optimal binning method and bins.

These findings align with broader research in CSCW and HCI on decision-making in both collaborative and individual contexts, where factors like cognitive load, task complexity, and domain expertise significantly influence workflows~\cite{roth2002decision, carroll1997human}. Developing tools that streamline decision-making could be particularly beneficial in collaborative environments, such as collaborative sensemaking~\cite{paul2010understanding, mahyar2014supporting} and group decision-making~\cite{owen2015collaborative}, especially for (bin) selection tasks~\cite{olson1997decision}. By integrating technology into these workflows, we can reduce cognitive load and offer decision support~\cite{turban2011decision, power2002decision}. Such decision support systems (DSS) have the potential to enhance analytical capabilities, streamline processes, and provide real-time feedback for tasks like mapmaking~\cite{power2002decision}. However, reliance on such systems requires appropriate training to avoid overdependence or misapplication~\cite{orlikowski1992learning}.

\paragraphHeadingSpace\bpstart{Detecting and mitigating less `objective' decision-making methods.} 
Our study revealed that peoples' mood on a particular day can influence their mapping decisions. While this finding can be perceived as anecdotal, it aligns with prior work describing how an individual's emotion and mood can negatively influence their job performance and satisfaction~\cite{cote1999affect} and impact the organization's processes and outcomes~\cite{maitlis2013sensemaking}. 
Yet, this remains an understudied area in CSCW and HCI, especially in collaborative organizational settings. 
For instance, how do emotions and moods of multiple collaborators influence collaborative processes and outcomes? How does the mode of collaboration (in-person versus remote work) factor in? How can these emotions and moods be efficiently detected and their impact subsequently mitigated?
These questions impact not just mapmaking but diverse domains, opening up new avenues for studying emotional dynamics during informal communication at work~\cite{kraut1990informal} and coordination in distributed organizational settings~\cite{srikanth2007coordination, fiore2003distributed}.

In addition to emotional influences, we also found that personal preferences (or rules of thumb) play a crucial role in the mapping outcome. 
Research on decision-making and cognitive biases highlights how individual differences in preferences can affect choices and judgments~\cite{tversky1974judgment, simon1955behavioral}.
Future work could investigate the design of tools that detect, warn about, and mitigate such `less objective' decision-making methods~\cite{kahneman2011thinking, lerner2015emotion}.

In a collaborative work setting, conflicts may arise when one individual's preferences clash with another's or with established cartographic principles. Effective collaboration in such cases is crucial for maintaining workflow efficiency and creativity. Prior research in CSCW has shown that personal preferences, including work style, creativity, and problem-solving approaches, can significantly impact team dynamics and performance. Misalignments in team members' preferences can lead to conflicts, inefficiencies, or creative tensions, hindering collaboration and productivity~\cite{amabile1988model}. To address these challenges, adaptive interfaces can be designed to cater to individual preferences, allowing users to work with tools they are most comfortable with while ensuring seamless integration into the larger team workflows~\cite{jameson2007adaptive, fischer2001user}. Additionally, tools can mediate conflicts adaptively by reconciling individual preferences with shared team goals, drawing from groupware and adaptive systems~\cite{grudin1994groupware}. By implementing such adaptive solutions, teams can not only improve collaboration but also achieve more successful project outcomes, balancing individual needs with collective goals.

\subsection{Facilitating Effective Interdisciplinary Collaborations in Organizations}
\paragraphHeadingSpace\bpstart{Minimizing friction due to interdisciplinary collaborations.}
Like other domains~\cite{kandel2012enterprise,muller2019data,walny2019data}, our study revealed that mapmaking is an interdisciplinary collaborative process involving mapmakers, subject matter experts, reviewers, designers, content-writers, publicists, across organizational hierarchies, even external clients. 
Even though the data and tools are readily available, a mapmaker does not generally work in isolation, underscoring the importance of shared experience and expertise to jointly define problems, share viewpoints, and create integrated knowledge~\cite{balakrishnan2011research, godemann2008knowledge, hammer2001enhancing, petts2008crossing, schmidt1992taking, stock2011defining, tress2007analysis}. 
However, we did notice two kinds of `friction' that negatively impact the collaboration process and/or its outcome~\cite{edwards2011science}: (1) cartographic principles are in conflict with mapmakers' personal preferences and organizational protocols; and (2) mapmakers used a wide set of terminologies and used terms interchangeably, e.g., choropleth maps were referred to as ``choropleths'', ``chlorofills'', ``thematic maps'', and ``colored maps''; similarly, bins were also referred to as ``classes'', ``categories'', ``buckets'', ``schemas'', ``algorithm outputs'', and ``ranges''.
Mapmaking can learn from prior CSCW and HCI studies that have documented evidence of such `friction' due to misaligned workflows~\cite{balakrishnan2011research}, non-standard data formats or differences in terminology~\cite{panagiotidou2022designing}, including human factors such as interpersonal relationships~\cite{kraut1986relationships, haythornthwaite2006learning}, competition~\cite{drago1991competition}, and biases~\cite{heilman2012gender}. 
For example, `friction' can be minimized by developing team-first tools and frameworks that streamline processes, foster shared understanding via standardization and shared vocabulary~\cite{edwards2011science, stock2011defining}, and address human factors.

\paragraphHeadingSpace\bpstart{Overcoming technological and organizational frustrations.} 
Mapmakers often require a combination of tools but often, they face frustrations with the tools provided by their organizations, as these tools may not always align with community standards, sufficiently support their mapping requirements, or fall within their areas of expertise or prior experience.
For instance, one of our interviewees desired a feature, but it was proprietary and not included in their organization's tool subscription, with no open-source alternative available; another mapmaker did not like the styling and font options available in their tool.
Not only technology, even organizational protocols and guidelines that are designed to ensure the replicability, consistency, transparency, and accountability of maps, can sometimes conflict with cartographic principles, putting mapmakers in a challenging position.

Such bureaucratic influences are not unique to mapmaking and have been shown to be barriers to technological adoption and innovation~\cite{obermeyer1990bureaucratic} across data-driven disciplines~\cite{obermeyer1990bureaucratic, kandel2012enterprise}.
One way to address these frustrations is a dual approach: improving existing tools or building open-source alternatives, and educating organizations to more closely understand the needs of the mapmakers and the tools available in the market.
It is also important for both mapmakers and organizations to navigate and balance the complexities between cartographic and organizational requirements~\cite{obermeyer1990bureaucratic}, e.g., develop expertise across multiple tools and as \expert{6} said, \emph{``move away from ``rules and policies'' to ``best practices'' as it is not always possible to follow [the former].''}
Lastly, integrating collaborative visualization systems~\cite{heer2008supporting} or leveraging participatory design approaches~\cite{muller1993participatory} could further enhance tool usability and alignment with user needs.

\subsection{Bridging the Disconnect between Theory, Research, and Practice}
\paragraphHeadingSpace\bpstart{Translating cartographic principles into practice.} Many mapmakers mentioned not \emph{exactly} following an `objective' cartographic process. For example, some did not analyze the data before mapping, some were unaware of certain binning methods or always used their `go to' method (even if it was unsuitable for the data). Many mapmakers, who did not have access to desired proprietary features in tools longed for more open-source alternatives.
This disconnect echoes broader concerns in CSCW and HCI about translating theoretical principles into practice~\cite{anya2015bridge, ackerman2000intellectual}. HCI researchers could help bridge these gaps and enhance adoption by facilitating a more seamless translation of `textbook' principles into practice, e.g., curate a repository of cartographic guidelines for reference, design interactive tools (e.g. ColorBrewer~\cite{brewer1994color}) that apply these guidelines and seamlessly integrate into mapmakers' workflows, and share mapping-related `success stories' for inspiration, and integrating them into GIS curricula.

\paragraphHeadingSpace\bpstart{Incorporating feedback channels from end-users.} To our surprise, none of our interviewee mapmakers mentioned providing a means for the audience of the published map to provide feedback; the only feedback provided is from teammates or clients.
This oversight means missing out on important insights that could improve how the content (map) is actually read and understood by their intended audience~\cite{lee2017news, tadelis2016reputation}. 
User-centered design processes highlight that end-user feedback can identify and fix usability issues early on, enhance visual information presentation, and ensure that maps meet the diverse needs of different users~\cite{norman1986user, norman1988psychology}.
For example, interactive online maps can allow viewers to post comments, submit a feedback form, write an email response; offline maps can solicit feedback via postal mail, email, telephone, or other viable communication modes.
Mapmakers could also capture indirect feedback by logging users' interactions with the interactive maps to study their behavior (e.g., how they pan, zoom, or apply filters).
By integrating these feedback channels, mapmakers can bring users directly into the mapping loop, resulting in more effective and user-friendly maps while fostering effective audience engagement and supporting collaborative and user-driven mapmaking processes.

\subsection{Designing Inclusive and Context-Aware Maps}

\paragraphHeadingSpace\bpstart{Embracing geo-political considerations in mapmaking.} Organizations that serve global audiences often face significant political constraints related to administrative borders, influencing the types of maps they create and the information they disclose. 
This issue, while not exclusive to information visualization, is particularly prominent in it, especially mapmaking. 
Factors like territorial disputes and geopolitical tensions restrict them from depicting certain boundaries or presenting data in potentially controversial ways. 
These constraints can skew geographical representations, challenging mapmakers to balance transparency with compliance to regulatory or diplomatic protocols, affecting the accuracy and completeness of maps. 
Addressing these challenges requires expertise in mapmaking as well  an understanding of international relations and local sensitivities. 
Providing training or education on international relations and geopolitics could better equip mapmakers to navigate these complexities effectively, ensuring maps remain informative, impartial, and relevant to diverse audiences.

\paragraphHeadingSpace\bpstart{Designing for everyone.} Most mapmakers chose colorblind-, web-, print-friendly color schemes to make their maps accessible for the target audience, media, and format; and when this is not possible, they resorted to other techniques, such as hatching or annotations. 
Mapmakers also used colors based on general principles (e.g., the forest area is green) and with attention to potential implications (e.g., political party colors) to make a map relatable and understandable; however, these can also have confusing implications (e.g., reds and blues have political implications in the U.S. or blue and pink may be associated with gender) and hence care must be taken. 
Most mapmakers had the reader in mind when choosing round numbers (e.g., 1,000--2,000) to reduce the cognitive burden on map readers, although different methods are chosen where statistical accuracy and precision is warranted.
These efforts illustrate how mapmakers try to design for everyone, reinforcing recent HCI efforts to publicize~\cite{smith2020disseminating}, democratize~\cite{corticelli2009human}, increase access~\cite{keates2000towards} to knowledge.
In the cases when they are unable to do so, they try to balance the needs and capabilities of the target audience and other metrics pertaining to cartographic and visual design principles, a practice which is also in adherence with HCI research on inclusive design guidelines~\cite{keates2002countering}. These principles can guide the development of universally accessible maps while addressing cross-cultural sensitivities.
Moreover, mapmakers also involve stakeholders other than users (e.g., clients, public relations experts, subject matter experts) during mapmaking, a practice followed by assistive technology experts as part of their inclusive design processes~\cite{herriott2014inclusive}. 
Future mapping efforts could consider prototyping alternative designs and conducting verification testing with end users before finalizing the map~\cite{herriott2014inclusive}.

\paragraphHeadingSpace\bpstart{Advocating for more such `behind the scenes' tours.}
Just because the map product is out there and widespread does not mean we know how it is made.
By dissecting a mapmaker's decision-making processes -- from data selection and binning to color choice and final styling -- researchers, practitioners, and educators can improve cartographic design guidelines and tooling, enhance educational curricula, and indirectly ensure credibility and accountability of the decisions taken using the maps. 
Not just mapmaking, undertaking such `behind the scenes' tours in other fields such as journalism and social media network analysis, can also present numerous opportunities.
For instance, interviewing content moderators can inform better content moderation strategies, enhance user privacy protections, and promote healthier online interactions, contributing to improved platform functionalities and broader discussions on digital ethics and societal impacts.
While such efforts may face challenges such as organizational confidentiality concerns, addressing these through anonymized data and clear ethical guidelines can ensure their feasibility and effectiveness.
Even researchers face challenges in recruiting participants, navigating scheduling conflicts, and reacting during contingencies, all of which requires patience, perseverance, and careful planning.
However, these efforts can eventually bridge gaps between academia and industry, fostering innovation and shared best practices.

%% file: src/sections/07-limitations.tex
There were a few limitations to our study. 
First, our sample of 13 organizations was selected using purposive sampling~\cite{patton2014qualitative} to capture a range of perspectives relevant to our primary research focus -- understanding how not-for-profit government organizations, federal agencies, and NGOs make (choropleth) maps. While this sample size may appear limited, it aligns with established qualitative research practices that emphasize achieving empirical saturation -- the point at which additional interviews were likely to elicit insights that had already been noted~\cite{guest2006how, morse2000determining}. This approach ensures depth and richness of data over breadth, which is particularly suitable for qualitative interview studies. 
Future work may interview a wider sample, including private firms and independent mapmakers (e.g., data journalists) for alternate perspectives.

Next, our sample of organizations was primarily located in the U.S. and India; however, many of these organizations not only produced maps for their respective countries but also supported international efforts through overseas collaborations and teams, addressing diverse mapping needs globally.
This global operational footprint ensured a broader perspective on mapping practices, supporting our belief that the insights captured were expansive enough to address our research objectives. Future work can explicitly include organizations based in additional regions to further explore how geographic context might influence specific mapping practices.

With respect to mapmaking specifically, despite our stated protocol to keep respondents' identities confidential, our mapmakers may not have revealed all tools, processes they use at their organization out of confidentiality concerns. 
Next, although we probed our interviewees about collaboration during mapmaking, we did not analyze how organizational hierarchies and team dynamics, including gender and diversity considerations impact this process; this may be a future work topic. 
Lastly, while our study was focused on univariate choropleth maps, future work should focus on multivariate choropleth maps and also other thematic map types such as graduated symbol maps and isolines (see \cite{field2021thematic}).

Regarding generalizability of our findings, they may be directly transferrable to those domains that have similar cultural, organizational, and geographic context as those of our interviewee organizations. 
For instance, research in CSCW has shown that context-specific factors -- such as organizational culture, work practices, and local norms -- play a significant role in shaping the adoption and effectiveness of collaborative technologies~\cite{ackerman2000intellectual, schmidt1992taking}. These factors can either enable or hinder the application of research findings in new settings. Readers are, hence, encouraged to critically assess these factors, recognizing that slight contextual misalignments could require adaptation~\cite{lincoln1985naturalistic}. By considering these contextual nuances, our work can still provide a robust foundation for advancing collaboration and technology use in diverse domains.

\section{Conclusion}
We interviewed 16 mapmakers from 13 well-known globally-serving government organizations, NGOs, and federal agencies about their choropleth mapmaking workflows. Our goal was to learn about how trained cartographers work within the provisions and structures of organizations.
We found that mapmaking is a complex undertaking involving interdisciplinary collaborations within and across organizations. While there are agreed-upon dicta, some processes vary across organizations due to other human factors (e.g., a mapmaker's personal preferences and `mood' on a given day) and external factors (e.g., requiring that choropleth maps use an organization's brand colors) that deviate from academic cartographic design principles. 
Our questions about binning and color choices resulted in relatively lively conversations with mapmakers. In no case did the mapmakers give short, concise answers to these questions (in our perspective), nor did they query why we were interested in these subjects; they seemed eager to discuss these `details'. 
This enthusiasm may suggest that these choices are central to choropleth mapping and agency in mapmaking, and should be further explored.
This grey area can be valuable for CSCW and HCI researchers and cartographers as they balance the territory between standards and practical graphic production in the future.